# Improvement of primary power standard through international comparison feedback

LUCIANO BRUNETTI, LUCA OBERTO[(1)], MARCO SELLONE
ISTITUTO NAZIONALE DI RICERCA METROLOGICA (INRIM) – STRADA DELLE CACCE, 91 – 10135, TORINO, ITALY
[(1)]CORRESPONDING AUTHOR: E-MAIL: L.OBERTO@INRIM.IT, TEL: + 39 (0)11 3919327, FAX: +39 (0)11 346384

*Abstract* - The high frequency primary power standard is univocally realized by means of a coaxial microcalorimeter, at least up to 40 GHz. The coaxial microcalorimeter is a broadband measurement system adjusted for effective efficiency measurement of power sensors both of bolometric and thermoelectric type. The critical point in the power standard realization is in the determination of the calibration constant $g$ of the microcalorimeter, a frequency dependent parameter that has huge impact on the accuracy assessment of the standard. The paper proposes a simple but powerful way for improving this accuracy by using the reference values provided by international key-comparisons which the microcalorimeter was involved in.

## I. Introduction

The microcalorimeter is a measurement system developed by National Measurement Laboratories (NMIs) to realize the primary power standards in the high frequency (HF) field, typically above 10 MHz. No commercial replica exists, of this instrument, that allows the measurement of the parasitic losses versus frequency of a power sensor, establishing in this way its effective efficiency $\eta_e$. When such a parameter is known, the power sensor becomes eligible for primary transfer standards while, in a wide sense, microcalorimeter is conventionally identified as the primary power standard [1]. The power sensor suitable for the microcalorimeter technique must be of thermal type. In the past it was common to use bolometers, but now thermoelectric power sensors appear as a more interesting alternative to the classical bolometric devices. In fact, an international power comparison was specifically adjusted to use thermoelectric travelling standards [2].

The Istituto Nazionale di Ricerca Metrologica (INRIM, formerly IEN "Galileo Ferraris"), was involved in this international exercise, classified as Key Comparison CCEM.RF-K10.CL, with a new twin type coaxial microcalorimeter fitted with 3.5 mm feeding lines [3]. Such system was suitable for effective efficiency measurement with a claimed uncertainty less than ±2% (with a coverage factor $k$=1) up to 26 GHz, that was the highest frequency also established by the comparison [2, 3].

Even though the performances of the INRIM coaxial microcalorimeter have been strongly improved in term of frequency band and accuracy by means of hardware and software updates, the results obtained in the CCEM.RF-K10.CL Comparison still are conditioning the INRIM Common Measurement Capabilities (CMCs) concerning the HF power measurement and standard. No other wide international exercise has been performed, in-deed, to prove and validate the actual performance of the INRIM coaxial microcalorimeter.

Actually, the INRIM twin coaxial microcalorimeter has been adjusted to calibrate thermoelectric power sensors, for which, e.g., we claim an uncertainty level less ±0.5% on a 26 GHz frequency band [4].

Anyway, independently of the improvements we have introduced to our measurement system [4, 5] and of the reliability test they must undergo, INRIM power measurement capability could be updated by using the same official results of the CCEM.RF-K10.CL comparison.

In this paper we propose a simple method to improve the quality of the primary power standard and consequently the INRIM CMCs, without additional measurement and using only the international comparison feedback.

## II. Primary power standard theory

The mathematical model that allows the effective efficiency determination of thermoelectric power sensors together with the microcalorimeter technique has been widely described by the same authors in the literature [3-7], therefore, in this section, we give only some fundamentals to provide the reader with enough background. All mathematical models proposed by us derive from the superimposition principle of the linear effects invoked together with the principle of equivalence of the thermal effects associated to a DC power that substitute a HF power [1].

For an adiabatic calorimeter, as the INRIM system schematized in Fig. 1, the response $e$ of the system electrical thermometer to power injection is given by:

$$e = \alpha \Delta T = \alpha R \left( K_A P_S + K_B P_L \right) \qquad (1)$$

where $\alpha$ is the Seebeck coefficient of the thermometer, $R$, a conversion constant depending on thermodynamic parameters of the microcalorimeter thermal load, $K_A$ and $K_B$, coefficients that describe the power separation between thermal load (i.e. the power sensor under calibration) and the feeding line, while $P_S$, $P_L$ are the power dissipated in the sensor and in the feeding line, respectively.

Assuming as *effective efficiency* ($\eta_e$) definition of a thermoelectric sensor, the ratio of the measured power $P_M$, i.e. the power converted in the DC output $U$ of the power sensor, to the total power absorbed by the same sensor $P_S = (P_M + P_X)$, that is [7]:

$$\eta_e = \frac{P_M}{P_M + P_X} \qquad (2)$$

where $P_X$ is the power loss in the sensor mount, then an expression of $\eta_e$ is:

$$\eta_e = g\frac{e_2}{e_1} \qquad (3)$$

in which $e_1$ and $e_2$ are DC-voltages corresponding to the two different thermodynamic equilibrium states of the system and related to the power injected into the system at two different frequencies, [3], while $g$ is a parameter that accounts for the microcalorimeter losses, basically located on the feeding line of the measurement channel, [6, 7].

For an ideal microcalorimeter, Eq. (3) reduces to a voltage ratio measurable with a typical relative uncertainty of units in $10^4$ and this could be considered the actual accuracy limit in the HF-power standard realization. In practice, the $g$-value can be slightly different from the unity, particularly at high frequencies, therefore its evaluation is mandatory to avoid accuracy degradation in the effective efficiency measurement.

At first glance, the $g$-determination can be done reversing (3), if a calorimetric load, i.e. a power sensor, of known or calculable effective efficiency is available. It has been demonstrated [7] that the best calculable calorimetric load is realized by short-circuiting the input of the same power sensor under calibration. In practice this operation is all but simple and free of risks. Indeed the substitution of the power sensor that normally behaves as matched load with a completely reflective one is not perfectly equivalent to an ideal short-circuit, neither from the electrical point of view nor from the thermal one. Alternative models to Eq. (3), that try to account for this could be proposed but, to our knowledge, the problem of *g*-determination still remains not enough solved. Furthermore, the hardware transformation required by the inversion of Eq. (3) can change significantly the thermodynamic equilibrium of the system, causing the introduction of bias that prevents a correct evaluation. This inconvenient could explain the discrepancies that sometimes appear in primary power standard realizations.

Anyway, Eq. (3) may be advantageously reversed when the reference value of a relevant Key Comparison is available. This means that every laboratory using the reference data of a comparison to check its system can expect to improve really its primary power standard. In this way, laboratories that participated to the comparison with a microcalorimeter can verify the goodness of their own calibration system and introduce the appropriate correction, if any. As each participant to a high level comparison normally uses a different microcalorimeter, there is a weak probability to collect data with the same bias errors.

In the next section we present an application of this concept to the INRIM HF-power standard by using the results of the Key Comparison CCEM.RF-K10.CL [2].

## III. Data analysis

Although in a HF power key comparison the significant measurand is $\eta_e$, CCEM.RF-K10.CL protocol asked the participants to provide the measurement result in term of calibration factor *K*. It is related to the effective efficiency by the following equation:

$$K = (1-|\Gamma|^2)\eta_e = (1-|\Gamma|^2)g\frac{e_2}{e_1} \qquad (4)$$

where $\Gamma$ is the reflection coefficient of the travelling standard, while the other quantities are the same previously defined. The aim of this request is to allow the participation also of NMIs that cannot operate a microcalorimeter. For such laboratories, the evaluation of $\eta_e$ may be only indirect and *K* is the quantity really measured [8].

Because for thermoelectric power sensors, an *effective efficiency* definition still does not exist that is unconditionally accepted and reported in the technical literature as for the bolometric case [9], the protocol of the comparison [2] suggested a specific definition, so to avoid misunderstanding and to assure uniformity among the results. Shortly, $\eta_e$ was defined as the ratio between a 1 kHz reference power and the HF power that produces the same sensor output. It is anyway easy to demonstrate that this definition is equivalent to the more intuitive one given by Eq. (2), [7]. The suggestion to perform a power substitution at 1 kHz instead of DC is to avoid the inclusion in the error budget of the effects of the contact thermo-voltages. Furthermore, the technical protocol suggested to introduce two corrective terms at voltage ratio $e_2/e_1$ to uniform the uncertainty evaluation among the participants. In the following we maintain the same formalism and conventions of the technical protocol; therefore, the mathematical model that we have to consider t is:

$$K = (1- |\Gamma|^2) g \left( \frac{e_2}{e_1} + \delta e_{th} + \delta e_U \right) \qquad (5)$$

where the term $\delta e_{th}$ is a correction term specifically aimed to consider system thermal drift and limited resolution of the system thermometer, while $\delta e_U$ is related to an imperfect power substitution between the reference (1 kHz) and calibration frequency. It is important to highlight that, although all the terms of Eq. (5) give an error contribution, only $g$ is significantly dominant. Therefore this is the term that we want to re-evaluate by using the feedback of the CCEM.RF-K10.CL.

According to the data reported in the final official report of the key comparison, the situation of the INRIM power standard is schematized in Table 1.

The reference values in the KCVR–$K$ column have been calculated considering only the results of the participants having an independent power standard, i.e. the microcalorimeter. These results were elaborated according to the documents reported in the references [10,11], to exclude outliers from the averaging process. Numbers with gray background have not been used to calculate the reference values because resulted outliers. Anyway by using the reference values KCVR–$K$ and the following equation set:

$$g = \frac{K}{(1-|\Gamma|^2)(e_R + \delta e_{th} + \delta e_U)}$$

$$\frac{\partial g}{\partial K} = \frac{1}{(1-|\Gamma|^2)(e_R + \delta e_{th} + \delta e_U)}$$

$$\frac{\partial g}{\partial \Gamma} = \frac{-2|\Gamma|K}{(1-|\Gamma|^2)^2 (e_R + \delta e_{th} + \delta e_U)} \tag{6}$$

$$\frac{\partial g}{\partial e_R} = \frac{\partial g}{\partial \delta e_{th}} = \frac{\partial g}{\partial \delta e_U} = \frac{-K}{(1-|\Gamma|^2)(e_R + \delta e_{th} + \delta e_U)^2}$$

we can update the values of the calibration constant $g$ of the INRIM (IEN) microcalorimeter, together with the associated uncertainty, according to the same criteria given in [11]. Calculations have been done assuming a Gaussian propagation of the uncertainty. Moreover, at the frequencies where the INRIM data concurred to define the KCRVs (figures reported with white background in column 4 and 5 of Table 1), a correlation exists so the appropriated terms of covariance have been included in the uncertainty evaluation [12] assuming an unitary correlation coefficient to be conservative. Table 2 reports the old $g$-value against the new ones. What is significant is the improvement of the $g$-value and, mainly, of its-uncertainty.

At this point, the new $g$-data can be used to calibrate another transfer standard by means of the microcalorimeter, generating so a better power reference. The operation is fully consistent at all the frequencies of the in-ternational comparison CCEM.RF-K10CL, provided the microcalorimeter inset, i.e. the feeding line hardware, has not been changed. The feeding line losses are, indeed, the main error source of the microcalorimeter and have the main influence on the $g$-value. The described process may be applied to check the validity of the accuracy improvement claimed by INRIM after the conclusion of the Key Comparison. In fact the modification that the microcalorimeter underwent so far did not interest the feeding lines but the thermostatization system, the power levelling loop, the calibration short circuit and the data analysis procedure. The results should be significant especially at the frequency for which INRIM was considered outlier (data showed with grey background in Table 1). At these frequencies there is no correlation between the KCRVs and the INRIM measurements.

## IV. Conclusions

We proposed a simple calculation process to generate a highly reliable HF power reference, at least to a fixed number of frequencies. The process considers the data feedback coming from an international power comparison and it should be useful to all the participating laboratories that use a microcalorimeter. The participants resulted outliers can utilize the reference data to generate a power standard removing, in their measurements, the undetected bias present at the moment of the international exercise. For the other laboratories, the feedback can be useful to validate and also reduce their claimed uncertainty. INRIM is in both situations. Actually our uncertainty on $g$ is less than ± 0.8% up to 26 GHz that, compared to an initial ±1.7%, demonstrates that we were able to remove the most significant error sources in the microcalorimeter calibration. If, in future power standard realizations, we will use the result of this calculation instead of performing an independent system calibration, we expect to obtain a standard with an uncertainty of the same order of magnitude because $g$ is the main error source.

Figure 1:

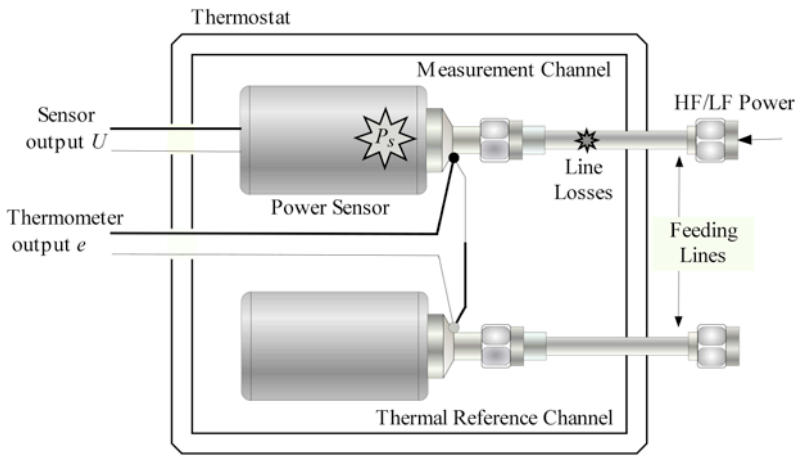

Figure caption:

Figure 1: *Scheme of the fundamental components of twin microcalorimeter [7].*

Table 1:

| Freq. (GHz) | **KCVR - $K$** | u($K$) | IEN - $K$ | IEN - u($K$) |
|---|---|---|---|---|
| 0.05 | **0.9866** | 0.0023 | 0.9895 | 0.0052 |
| 1 | **0.9760** | 0.0027 | 0.9817 | 0.0063 |
| 10 | **0.9416** | 0.0014 | 0.9344 | 0.0135 |
| 18 | **0.9317** | 0.0034 | 0.9273 | 0.0138 |
| 20 | **0.9276** | 0.0041 | 0.9192 | 0.0131 |
| 23 | **0.9252** | 0.0037 | 0.9115 | 0.0190 |
| 26 | **0.9228** | 0.0038 | 0.9003 | 0.0187 |

Table 2:

| Freq. (GHz) | IEN - $g$ | IEN - u($g$) | new $g$ | new u($g$) |
|---|---|---|---|---|
| 0.05 | 1.0119 | 0.0061 | 1.0105 | 0.0034 |
| 1 | 1.0636 | 0.0060 | 1.0567 | 0.0058 |
| 10 | 1.1559 | 0.0170 | 1.1653 | 0.0028 |
| 18 | 1.2465 | 0.0160 | 1.2539 | 0.0097 |
| 20 | 1.2609 | 0.0160 | 1.2717 | 0.0067 |
| 23 | 1.2773 | 0.0230 | 1.2994 | 0.0058 |
| 26 | 1.3191 | 0.0230 | 1.3456 | 0.0062 |

Table captions:

Table 1: *Results of INRIM (IEN) measurements against the Key Comparison Reference Values (KCRVs) of the CCEM.RF–K10.CL Comparison.*

Table 2: *Microcalorimeter calibration factor g measured by INRIM (IEN) against the corrected value through the CCEM.RF-K10CL. Gray background means no correlation exists between KCRVs and INRIM (IEN) measurements.*